\documentclass{article}
\usepackage[nohead]{realA4}
\usepackage{enumerate}

\usepackage{epsfig}

\usepackage{amsthm,amssymb,amsmath,latexsym}

\theoremstyle{plain}

\newtheorem{lemma}{Lemma}

\theoremstyle{definition}

\numberwithin{equation}{section}

\title{\vspace*{-1.3cm} \bf Oblivious Transfer  and Quantum Non-Locality}

\author{\Large Stefan Wolf\,\footnote{D\'epartement d'Informatique et R.O., 
Universit\'e de Montr\'eal, Canada. Email: 
{wolf@iro.umontreal.ca}.}\ \ \ \ \ \ \ \ 
J\"urg Wullschleger\,\footnote{D\'epartement d'Informatique et R.O., 
Universit\'e de Montr\'eal, Canada. Email: 
{wullschj@iro.umontreal.ca}.}}

\date{}

\begin{document}

\maketitle

\vspace*{-0.7cm}
\begin{abstract}
\noindent
{\em Oblivious transfer}, a central functionality in modern
cryptography, allows a party to send two one-bit messages to another
who can choose one of them to read, remaining ignorant about the
other, whereas the sender does not learn the receiver's
choice. Oblivious transfer the security of which is
information-theoretic for both parties is known impossible to achieve
from scratch. | The joint behavior of certain bi-partite quantum
states is {\em non-local}, i.e., cannot be explained by shared
classical information. In order to better understand such behavior,
which is classically {\em explainable\/} only by communication, but
does not {\em allow\/} for it, Popescu and Rohrlich have described a
``non-locality machine'': Two parties both input a bit, and both get a
random output bit the XOR of which is the AND of the input bits. | We
show a close connection, in a cryptographic sense, between OT and the
``PR primitive.'' More specifically, unconditional OT can be achieved
from a single realization of PR, and {\em vice versa}. Our reductions,
which are single-copy, information-theoretic, and perfect, also lead
to a simple and optimal protocol allowing for inverting the direction
of OT.
\end{abstract}

\section{Introduction}

\subsection{Oblivious Transfer and Oblivious Keys}

{\em Oblivious transfer}~\cite{Rabin81}, {\em OT\/} for short, is a
functionality of great importance~\cite{Kilian88} in cryptography or,
more precisely, {\em secure two-party computation}, where two parties,
who mutually distrust each other, want to collaborate with the
objective of achieving a common goal, e.g., evaluate a function to
which both hold an input|but without revealing unnecessary information
about the latter. In (chosen one-out-of-two bit) OT, one of the
parties, the {\em sender}, inputs two bits $x_0$ and $x_1$, whereas
the other party has a {\em choice\/} bit $c$. The latter then learns
$x_c$, but remains ignorant about the other message bit $x_{1-c}$. The
sender, on the other hand, does not learn any information about $c$.

Various ways, based on public-key encryption, for instance, have been
proposed for realizing OT, where the security for one of the parties,
however, is only computational. In fact, oblivious transfer is
impossible to achieve in an unconditionally secure way for both
parties|even when they are connected by a quantum
channel~\cite{mayers}. On the other hand, it has been shown that
unconditionally secure OT can be reduced to weak information-theoretic
primitives such as simply a noisy communication
channel~\cite{Crepeau97},\, \cite{cmw04} or so-called {\em universal
OT\/}~\cite{bcw}.

A recent result~\cite{wwec04} shows that OT can be stored: Given one
realization of OT, a sample of distributed random variables $X$ (known
to $A$) and $Y$ (known to $B$) can be generated, where the joint
distribution $P_{XY}$ is such that $X$ and $Y$ can be used to realize
an instance of OT. We will call the distributed pair of random
variables $(X,Y)$ an {\em oblivious key\/} or OK for short; in some
sense, as we will see, this is the {\em local\/} (hidden-variable)
part of OT (as opposed to non-local systems and behavior, see
Section~1.2). Another consequence|observed in~\cite{wwec04}|is that,
since OK is symmetric, OT is, too. This solved a long-standing open
problem posed in~\cite{cs}.

\subsection{Quantum Non-Locality and the Popescu-Rohrlich Primitive}

{\em Entangled\/} but possibly distant two-partite quantum systems can
show a joint behavior under measurements that cannot be explained by
``locality'' or hidden variables, i.e., distributed classical
information; such behavior is called {\em non-local}.  There exists,
for instance, a so-called {\em maximally entangled\/} state
$|\psi^{-}\rangle=(|01\rangle-|10\rangle)/\sqrt{2}$ with the following
properties.  If the parties $A$ and $B$ controlling the two parts of
the system both choose between two fixed possible bases for measuring
their system in (where this pair of bases is not the same for the two
parties), where the measurement outcome can be $0$ or $1$ in both
cases, then the following statistics are observed.  (Here, the two
possible bases for each party are called $0$ and $1$, too.)
\begin{eqnarray*}
p_{00} & := & {\rm Prob\, } [\, \mbox{outcome $A$}=\mbox{outcome $B$}\, |\, \mbox{basis $A$}=0\, ,\ \mbox{basis $B$}=0\, ]\ \ =\ \ 0\ ,\\
p_{01} & := & {\rm Prob\, } [\, \mbox{outcome $A$}=\mbox{outcome $B$}\, |\, \mbox{basis $A$}=0\, ,\ \mbox{basis $B$}=1\, ]\ \ =\ \ 1/4\ ,\\
p_{10} & := & {\rm Prob\, } [\, \mbox{outcome $A$}=\mbox{outcome $B$}\, |\, \mbox{basis $A$}=1\, ,\ \mbox{basis $B$}=0\, ]\ \ =\ \ 1/4\ ,\\
p_{11} & := & {\rm Prob\, } [\, \mbox{outcome $A$}=\mbox{outcome $B$}\, |\, \mbox{basis $A$}=1\, ,\ \mbox{basis $B$}=1\, ]\ \ =\ \ 3/4\ .
\end{eqnarray*}

\noindent
It has been shown that such statistics are impossible to achieve
between two parties who cannot communicate when they share arbitrary
{\em classical\/} information only (i.e., agree on a classical
strategy beforehand). More precisely, the so-called CHSH {\em Bell
inequality\/} is violated since
\[
p_{00}+p_{01}+p_{10}<p_{11}
\]
holds. It is, on the other hand, important to note that this non-local
behavior is ``weaker'' that communication between $A$ and $B$ and does
not allow for such|fortunately, since such a possibility would be in
contradiction with relativity.

With the objective of achieving a better understanding of such
``non-local behavior,'' Popescu and Rohrlich~\cite{pr} defined a
``non-locality primitive'' behaving in a similar way, but where the
probabilities $p_{ij}$ are
\[
p_{00}=p_{01}=p_{10}=0\, ,\ \ p_{11}=1\ .
\]
In other words, both parties have an input bit (corresponding to the choice of the  
basis in the quantum model) $U$ and $V$ and get an output bit 
$X$ and $Y$, respectively, where $X$ and $Y$ are random bits satisfying 
\[
X\oplus Y=U\cdot V= U \mbox{ AND }V\ .
\]
It is important to note, however, that the behavior of this ``PR
primitive'' cannot, although it does not allow for communication
either, be obtained from any quantum state|it violates a ``quantum
Bell inequality'' that is even valid for the behavior of quantum
states. On the other hand, the primitive {\em does\/} allow for
perfectly simulating the behavior of a maximally entangled quantum bit
pair under {\em any\/} possible measurement~\cite{nlm}. The latter has
been shown possible, for instance, also between parties who may
communicate {\em one\/} classical bit~\cite{bt}\, \cite{bct}, but the
possibility of achieving the same with the PR primitive is of
particular interest since this functionality does not allow for any
communication.

\subsection{Two-Party Information-Theoretic Primitives and Reductions}

The three {\em information-theoretic primitives\/} or two-party
functionalities described in the previous sections can be modeled by
their mutual input-output behavior, i.e., by a conditional probability
distribution $P_{XY|UV}$, where $U$, $V$, $X$, and $Y$ are the two
parties' input and output, respectively (see Figure~1).

\begin{figure}[h]
\begin{center}

\begin{picture}(0,0)%
\includegraphics{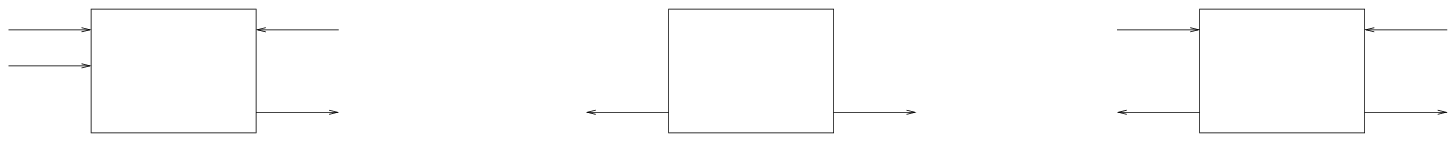}%
\end{picture}%
\setlength{\unitlength}{1302sp}%
\begingroup\makeatletter\ifx\SetFigFont\undefined%
\gdef\SetFigFont#1#2#3#4#5{%
  \reset@font\fontsize{#1}{#2pt}%
  \fontfamily{#3}\fontseries{#4}\fontshape{#5}%
  \selectfont}%
\fi\endgroup%
\begin{picture}(22815,4260)(376,-3994)
\put(16651,-2386){\makebox(0,0)[lb]{\smash{{\SetFigFont{8}{9.6}{\familydefault}{\mddefault}{\updefault}$a$}}}}
\put(16651,-1186){\makebox(0,0)[lb]{\smash{{\SetFigFont{8}{9.6}{\familydefault}{\mddefault}{\updefault}$x$}}}}
\put(16501, 14){\makebox(0,0)[lb]{\smash{{\SetFigFont{8}{9.6}{\familydefault}{\mddefault}{\updefault}$A$}}}}
\put(22351,-1186){\makebox(0,0)[lb]{\smash{{\SetFigFont{8}{9.6}{\familydefault}{\mddefault}{\updefault}$y$}}}}
\put(22351,-2386){\makebox(0,0)[lb]{\smash{{\SetFigFont{8}{9.6}{\familydefault}{\mddefault}{\updefault}$b$}}}}
\put(17776,-3886){\makebox(0,0)[lb]{\smash{{\SetFigFont{8}{9.6}{\familydefault}{\mddefault}{\updefault}where $a\oplus b =x\cdot y$}}}}
\put(19276,-1786){\makebox(0,0)[lb]{\smash{{\SetFigFont{8}{9.6}{\familydefault}{\mddefault}{\updefault}$PR$}}}}
\put(22576, 14){\makebox(0,0)[lb]{\smash{{\SetFigFont{8}{9.6}{\familydefault}{\mddefault}{\updefault}$B$}}}}
\put(8776, 14){\makebox(0,0)[lb]{\smash{{\SetFigFont{8}{9.6}{\familydefault}{\mddefault}{\updefault}$A$}}}}
\put(6226,-1186){\makebox(0,0)[lb]{\smash{{\SetFigFont{8}{9.6}{\familydefault}{\mddefault}{\updefault}$c$}}}}
\put(6226,-2386){\makebox(0,0)[lb]{\smash{{\SetFigFont{8}{9.6}{\familydefault}{\mddefault}{\updefault}$x_c$}}}}
\put(526,-1711){\makebox(0,0)[lb]{\smash{{\SetFigFont{8}{9.6}{\familydefault}{\mddefault}{\updefault}$x_1$}}}}
\put(526,-1186){\makebox(0,0)[lb]{\smash{{\SetFigFont{8}{9.6}{\familydefault}{\mddefault}{\updefault}$x_0$}}}}
\put(376, 14){\makebox(0,0)[lb]{\smash{{\SetFigFont{8}{9.6}{\familydefault}{\mddefault}{\updefault}$A$}}}}
\put(10726,-3661){\makebox(0,0)[lb]{\smash{{\SetFigFont{8}{9.6}{\familydefault}{\mddefault}{\updefault}where $Y=X_C$}}}}
\put(3151,-1786){\makebox(0,0)[lb]{\smash{{\SetFigFont{8}{9.6}{\familydefault}{\mddefault}{\updefault}$OT$}}}}
\put(11626,-1786){\makebox(0,0)[lb]{\smash{{\SetFigFont{8}{9.6}{\familydefault}{\mddefault}{\updefault}$OK$}}}}
\put(8026,-2386){\makebox(0,0)[lb]{\smash{{\SetFigFont{8}{9.6}{\familydefault}{\mddefault}{\updefault}$X_0,X_1$}}}}
\put(14551,-2386){\makebox(0,0)[lb]{\smash{{\SetFigFont{8}{9.6}{\familydefault}{\mddefault}{\updefault}$C,Y$}}}}
\put(6451, 14){\makebox(0,0)[lb]{\smash{{\SetFigFont{8}{9.6}{\familydefault}{\mddefault}{\updefault}$B$}}}}
\put(14776, 14){\makebox(0,0)[lb]{\smash{{\SetFigFont{8}{9.6}{\familydefault}{\mddefault}{\updefault}$B$}}}}
\end{picture}%

\end{center}
\caption{Oblivious transfer, oblivious key, and the  Popescu-Rohrlich primitive.}
\label{otbox}
\end{figure}

\noindent
In Section~2, we will show simple perfect and single-copy
information-theoretic reductions between the three primitives|in some
sense, they are, provocatively speaking, all the same.

More precisely, a single-copy reduction of a primitive $P_2$ to
another $P_1$ means that the functionality $P_2$ can be realized given
one instance of $P_1$. Hereby, no computational assumptions have to be
made. {\em Perfect\/} means that no non-zero failure probability has
to be tolerated.

Note, however, that the reduction protocol may use communication; of
course, because from a ``communication and locality viewpoint,'' the
three primitives are very different: OT allows for communication, PR
does not, but is non-local, whereas OK is simply distributed classical
information, i.e., ``local.''

Although we keep an eye on this communication in the reductions|all
our reductions minimize the required amount of communication|, our
interest is {\em privacy\/}: When $P_2$ is obtained from $P_1$, say,
then both parties must not obtain more information than specified for
$P_2$. In other words, our viewpoint is the one of {\em
cryptography\/} rather than of communication-complexity theory.  Note
that our reductions have the property that a party who is misbehaving
in the protocol {\em cannot\/} obtain more information than specified
(but possibly violate the privacy of her proper inputs).

\section{Single-Copy Reductions Between OT, OK, and PR}

\subsection{PR from OT}

\begin{lemma} \label{lemma:OT2NL}
Using one instance of $OT$, we can simulate $PR$.
\end{lemma}

\begin{proof}
$B$ chooses $c = y$. $A$ chooses $a$ at random and sends $x_0 = a$
and $x_1 = x \oplus a$ with $OT$. $B$ receives $x_c$ and outputs $b =
x_c$. $A$ outputs $a$. We have $b = x_c = a \oplus x c = a \oplus xy$.
\end{proof}

\subsection{OK from PR}

\begin{lemma} \label{lemma:NL2OK}
Using one instance of $PR$, we can  simulate $OK$.
\end{lemma}

\begin{proof}
$A$ and $B$ choose $x$ and $y$ at random. $B$ outputs $C = y$ and $Y =
b$.  $A$ outputs $X_0 = a$ and $X_1 = a \oplus x$. We have $Y = b = xy
\oplus a = X_C$.
\end{proof}

\subsection{OK from OT}

\begin{lemma} \label{lemma:OT2OK}
Using one instance of $OT$, we can  simulate $OK$.
\end{lemma}

\begin{proof}
Follows directly from Lemmas \ref{lemma:OT2NL} and
\ref{lemma:NL2OK}. We get the following protocol: $A$ and $B$ choose
all their input at random.  $A$ outputs her inputs, $B$ his input and
his output.
\end{proof}

\subsection{OT from PR}

\begin{lemma} \label{lemma:NL2OT}
Using one instance of $PR$, we can  simulate $OT$ using one
bit of communication.
\end{lemma}

\begin{proof}
$A$ inputs $x = x_0 \oplus x_1$. $B$ inputs $y = c$. $A$ gets $a$
and $B$ gets $b$. $A$ sends $m=x_0 \oplus a$ to $B$. $B$ outputs $y
= m \oplus b$.  We have $y = m \oplus b = x_0 \oplus a \oplus b = x_0
\oplus (x_0 \oplus x_1)c = x_c$.

Since $A$ does not receive any message from $B$, she gets no
information about $c$. $B$ only receives one bit, which is equal to
$x_c$.
\end{proof}

\noindent
In $PR$, no communication takes place, but we are able to send one bit
using $OT$. Hence, at least one bit of communication is needed to
simulate $OT$ by $PR$.

\subsection{PR from OK}

\begin{lemma} \label{lemma:OK2NL}
Using one instance of $OK$, we can  simulate $PR$ using two
bits of communication.  
\end{lemma}

\begin{proof}
$A$ sends $m_a = x \oplus X_0 \oplus X_1$ to $B$. $B$ sends $m_b = y
\oplus C$ to $A$.  $A$ outputs $a = X_0 \oplus (X_0 \oplus X_1)m_b
+ m_a m_b$. $B$ outputs $b= Y \oplus C m_a$.  We have $X_0 \oplus Y =
(X_0 \oplus X_1)C$. Hence
\begin{eqnarray*}
a\oplus b
&=& X_0 \oplus (X_0 \oplus X_1)m_b \oplus m_a m_b \oplus Y \oplus C m_a \\
&=& m_a m_b \oplus (X_0 \oplus X_1)m_b \oplus C m_a \oplus (X_0 \oplus X_1)C \\
&=& (m_a \oplus X_0 \oplus X_1)(m_b \oplus C) = xy
\end{eqnarray*}
They both send their inputs ``XORed'' with $(X_0 \oplus X_1)$ and $C$,
respectively. Since the other party has no information about these
values, this is a one-time pad, and they receive no information about
the other's input.
\end{proof}

We show that the two bits of communication are optimal in this case.
Let us assume that there exists a protocol using only one-way
communication from $A$ to $B$. Since $B$ can calculate his output
$b_0$ for both inputs for $y=0$ and $b_1$ for $y=1$,  we have $a \oplus
b_0 \oplus a \oplus b_1 = x (1 \oplus 0)$, and, therefore, $b_0 \oplus
b_1 = x$. 

\subsection{OT from OK}

\begin{lemma}
Using one instance of $OK$, we can simulate $OT$ using three bits of
communication.
\end{lemma}

\begin{proof}
Follows directly from Lemmas~4 and~5. Alternatively, we can use the
BBCS protocol~\cite{BBCS92}, which requires three bits of
communication as well.  Here, $B$ sends $m = c \oplus C$ to $A$,
whereas $A$ sends $m_0 = x_0 \oplus X_{m}$ and $m_1 = x_1 \oplus X_{1
\oplus m}$ to $B$. $B$ outputs $y = m_c
\oplus Y$.  We have $y = m_c \oplus Y = x_c \oplus X_{c \oplus m}
\oplus Y = x_c \oplus X_{C} \oplus Y = x_c$.

$B$'s message does not give any information about $c$ to $A$, since it
is ``one-time padded'' with the value $C$ about which $A$ has no
information. $B$ knows either $X_0$ or $X_1$ but has no information
about the other value. So, either $x_0$ or $x_1$ gets ``one-time
padded,'' and $B$ obtains information about that value, even if he is
given the other value.
\end{proof}

Three bits of communication are optimal: First of all, two-way
communication is needed. If $A$ would send less than two bits, but
still in such a way that $B$ would get the bit he wants, then $A$
would have to know which bit $B$ has chosen.

\section{Optimally Reversing OT}

OT is {\em a priori\/} an asymmetric functionality, and the
possibility of inverting its orientation has been investigated, for
instance, in~\cite{cs}, where a protocol was given using $n$
realizations of OT from $B$ to $A$|called TO|in order to obtain one
realization from $A$ to $B$, where a failure probability exponentially
small in $n$ has to be tolerated. Since, however, PR is a symmetric
functionality, our reductions imply that OT is as well. More
precisely, the reductions of OT to PR and {\em vice versa\/} can be
put together to the following protocol inverting OT. This reduction of
OT to TO given in~\cite{wwec04}, is single-copy,
information-theoretic, perfect, and minimizes the required additional
communication.

\subsection{OT from TO}

\begin{lemma}
Using one instance of $TO$, we can simulate $OT$
using one bit of communication.
\end{lemma}

\begin{proof}
$A$ inputs $x_0 \oplus x_1$ to $TO$. $B$ chooses a random bit $r$ and
inputs $r$ and $r \oplus c$ to $TO$. $A$ receives $a$ and sends $m =
x_0 \oplus a$ to $B$. $B$ outputs $y = r \oplus m$.  We have $y = r
\oplus m = r \oplus x_0 \oplus r \oplus (x_0 \oplus x_1)c = x_c$.

$A$ does not get any message from $B$, so she does not get any
information about $c$. $B$ get one message by $A$, which is either
equal to $b_0$, if the XOR of his input values is $0$, and $b_1$
otherwise. If he does not choose $r$ at random, $A$ might be able to
get the value $c$, but there is no advantage for $B$.
\end{proof}

\noindent
The protocol is obviously optimal since $A$ can communicate one bit
with $B$ using $OT$|which she cannot using $TO$.

\subsection{OK from KO}
Finally, we show that an $OK$ can easily be reversed, without any
communication.

\begin{lemma}
Using one instance of $KO$, we can simulate $OK$
without any communication. 
\end{lemma}

\begin{proof}
$A$ gets $X_0$ and $X_1$ from $KO$, and $B$ gets $C$ and $Y$.
$A$ outputs $\overline C = X_0 \oplus X_1$ and $\overline Y = X_0$, and $B$
outputs $\overline X_0 = Y$ and $\overline X_1 = Y \oplus C$.

We have $Y = X_C$ and $\overline X_{\overline C}
= Y \oplus C (X_0 \oplus X_1)
= X_0 \oplus (X_0 \oplus X_1) C \oplus C (X_0 \oplus X_1) = X_0 = \overline Y$.
\end{proof}

\noindent
The $OK$ primitive can also be defined in a symmetric way: It is the
distribution that we get when both $A$ and $B$ input a random bit to
$PR$.

\section{Concluding Remarks}
We have shown a close connection between the important cryptographic
functionality of oblivious transfer and quantum non-locality, more
precisely, the ``non-locality machine'' of Popescu and Rohrlich: they
are, modulo a small amount of (classical) communication, the same|one
can be reduced to the other. As a by-product, we have obtained the
insight that {\em OT is symmetric\/}: One instance of OT from $B$ to
$A$ allows for the same functionality from $A$ to $B$ in a perfect
information-theoretic sense. Figure~2 shows the reductions between the
different functionalities discussed above. The (optimal) numbers of
bits to be communicated are indicated.

\begin{figure}[h]\label{chreis}
\hbox{
\centerline{
\psfig{figure=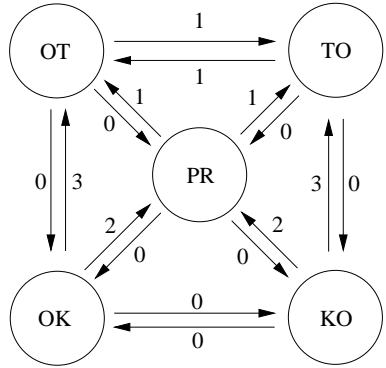,width=5cm}}}
\caption{The reductions between $OT$, $TO$, $PR$, $OK$,
and $KO$, and their communication costs. All reductions are prefect and optimal.}
\end{figure}

In has been shown in~\cite{nlm} that the behavior of an EPR pair can
be perfectly simulated without any communication if one realization of
the PR primitive is available. However, this reduction, although it
yields the correct statistics with respect to the two parts' behavior,
is not ``cryptographic'' or ``private'' in the sense of our
reductions: The parties are tolerated to obtain more information about
the other party's outcome than they would when actually measuring an
EPR pair. We state as an open problem to simulate, in this stronger
sense, the behavior of an EPR pair using the PR primitive.

\end{document}